\documentclass[journal]{IEEEtran}

\usepackage{scalerel}

\usepackage{graphicx}
\usepackage{amsfonts}
\usepackage{bm}
\usepackage[rgb]{xcolor}

\usepackage{multirow}
\usepackage{makecell}
\usepackage{booktabs}

\usepackage{hyperref}

\usepackage{subcaption}

\usepackage{enumitem} 
\usepackage{tabularx} 

\usepackage{algorithm}
\usepackage[noend]{algpseudocode}

\newcommand{\x} {\bm{x}}
\newcommand{\y} {\bm{y}}
\newcommand{\h} {\bm{h}}

\newcommand{\vecc} {\bm{c}}
\newcommand{\veca} {\bm{a}}
\newcommand{\DTTS} {\bm{D}_{\text{TTS}}}
\newcommand{\DASR} {\bm{D}_{\text{ASR}}}
\newcommand{\DVC} {\bm{D}_{\text{VC}}}
\newcommand{\Enc} {\text{Enc}}
\newcommand{\Dec} {\text{Dec}}
\newcommand{\head} {\text{head}}
\newcommand{\dmodel} {d_{\text{model}}}
\newcommand{\att} {\text{Att}}

\usepackage{cite}
\usepackage{amsmath}

\begin{document}

\title{Pretraining Techniques for Sequence-to-Sequence Voice Conversion}

\author{
	Wen-Chin~Huang,~\IEEEmembership{Student Member,~IEEE,}
	Tomoki Hayashi,
	Yi-Chiao Wu,~\IEEEmembership{Student Member,~IEEE,}
	Hirokazu Kameoka,~\IEEEmembership{Senior Member,~IEEE,}
	and~Tomoki Toda,~\IEEEmembership{Senior Member,~IEEE}

\thanks{Wen-Chin Huang and Yi-Chiao Wu are with the Graduate School of Informatics, Nagoya University, Japan. E-mail: wen.chinhuang@g.sp.m.is.nagoya-u.ac.jp, yichiao.wu@g.sp.m.is.nagoya-u.ac.jp}
\thanks{Tomoki Hayashi is with the Human Dataware Lab. Co. Ltd. and the Graduate School of Information Science, Nagoya University, Japan. E-mail: hayashi.tomoki@g.sp.m.is.nagoya-u.ac.jp}
\thanks{Hirokazu Kameoka is with the NTT Communication Science Laboratories, Nippon Telegraph and Telephone Corporation, Atsugi, Kanagawa, 243-0198 Japan. E-mail: kameoka.hirokazu@lab.ntt.co.jp}
\thanks{Tomoki Toda is with the Information Technology Center, Nagoya University, Japan. E-mail: tomoki@icts.nagoya-u.ac.jp}
}

\markboth{IEEE/ACM TRANSACTIONS ON AUDIO, SPEECH AND LANGUAGE PROCESSING}%
{Shell \MakeLowercase{\textit{et al.}}: Bare Demo of IEEEtran.cls for IEEE Journals}
%

\maketitle

\begin{abstract}
Sequence-to-sequence (seq2seq) voice conversion (VC) models are attractive owing to their ability to convert prosody. Nonetheless, without sufficient data, seq2seq VC models can suffer from unstable training and mispronunciation problems in the converted speech, thus far from practical. To tackle these shortcomings, we propose to transfer knowledge from other speech processing tasks where large-scale corpora are easily available, typically text-to-speech (TTS) and automatic speech recognition (ASR). We argue that VC models initialized with such pretrained ASR or TTS model parameters can generate effective hidden representations for high-fidelity, highly intelligible converted speech. We apply such techniques to recurrent neural network (RNN)-based and Transformer based models, and through systematical experiments, we demonstrate the effectiveness of the pretraining scheme and the superiority of Transformer based models over RNN-based models in terms of intelligibility, naturalness, and similarity.
\end{abstract}

\begin{IEEEkeywords}
voice conversion, sequence-to-sequence, pretraining, transformer
\end{IEEEkeywords}

\section{Introduction}

\IEEEPARstart{V}{oice} conversion (VC) aims to convert the speech from a source to that of a target without changing the linguistic content \cite{VC}. In conventional VC systems, with the assumption that a parallel corpus, i.e, pairs of speech samples with identical linguistic contents uttered by both source and target speakers, exists, an \textit{analysis\textemdash conversion \textemdash synthesis} paradigm is often adopted \cite{GMM-VC}. First, a high-quality vocoder such as WORLD \cite{WORLD} or STRAIGHT \cite{STRAIGHT} is utilized to extract different acoustic features, such as spectral features and fundamental frequency (F0). These features are converted separately, and a waveform synthesizer finally generates the converted waveform using the converted features. Past VC studies have focused on the conversion of spectral features while only applying a simple linear transformation to F0. In addition, the conversion is usually performed frame-by-frame, i.e, the converted speech and the source speech are always of the same length. This restricts the modeling of the speaking rate. To summarize, the conversion of prosody, including F0 and duration, is overly simplified in the past VC literature.

This is where sequence-to-sequence (seq2seq) models \cite{S2S} can play a role. Modern seq2seq models, often equipped with an attention mechanism \cite{S2S-NMT-Bah, S2S-NMT-Luong} to implicitly learn the alignment between the source and output sequences, can generate outputs of various lengths and capture long-term dependencies. This ability makes the seq2seq model a natural choice to convert prosody in VC. It has been shown that seq2seq VC models can outperform conventional frame-wise VC systems, especially in terms of conversion similarity. This is owing to the fact that the suprasegmental characteristics of F0 and duration patterns well handled in seq2seq VC models are closely correlated with the speaker identity.

Despite the promising results, seq2seq VC models suffer from two major problems. First, seq2seq models usually require a large amount of training data to generalize well, although parallel VC corpora are often hard to collect. Second, as pointed out in \cite{S2S-Text-VC}, the converted speech in seq2seq VC systems still suffers from mispronunciations and other linguistic-inconsistency problems, such as inserted, repeated and skipped phonemes. This is mainly due to the failure in attention alignment learning, which can also be seen as a consequence of insufficient data.

As one popular solution to limited training data, pretraining has been regaining attention in recent years, where knowledge from massive, out-of-domain data is transferred to aid learning in the target domain. This concept is usually realized by learning universal, high-level feature representations. In the field of computer vision, \textit{supervised} pretraining (e.g. ImageNet classification \cite{ILSVRC15, how-transferable, DeCAF}) followed by fine-tuning on tasks with less training data (e.g. object detection \cite{R-CNN, fast-R-CNN, faster-R-CNN}, segmentation \cite{U-Net, Mask-R-CNN} and style transfer \cite{style-transfer, perceptual-loss}) often leads to state-of-the-art results. On the other hand, many natural language understanding (NLU) tasks learn rich representation through an \textit{self-supervised} language model objective \cite{elmo, ULMFiT, openai-gpt, bert, openai-gpt-2}, which have also been shown to boost performance. Efforts have also been dedicated to speech representation learning for tasks including automatic speech recognition (ASR) \cite{wav2vec} and speaker identification/verification \cite{speaker-representation-MI}.

In our prior work \cite{VTN}, we proposed a text-to-speech (TTS) pretraining strategy for seq2seq VC. Instead of learning \textit{feature representations}, we focused on finding \textit{model parameters} that can easily learn effective representations in an \textit{supervised} manner. We initially adopted the text-to-speech (TTS) task based on the facts that (1) in a broader definition, both VC and TTS can be categorized into the speech synthesis task \cite{nautilus}, so the respective models should share some characteristics. (2) Vast studio-level large-scale TTS corpora have been made easily accessible by the community. It was demonstrated that the VC model initialized with TTS pretrained model parameters can generate high-quality, highly intelligible speech even with very limited training data.

In this work, we conducted complete experiments to fully analyze and understand this pretraining technique. In addition to TTS, we considered another task for knowledge transfer: ASR. As one of the most popular research fields in speech processing, ASR also uses $\langle text, speech \rangle$ data as TTS does, thus it is natural to consider ASR as an alternative task to transfer from. We provide systematical experimental results, and analysis of the hidden representation space learned with different pretraining tasks suggests the more effective source of knowledge transfer. We also examined two different model architectures: recurrent neural networks (RNNs) and Transformers, and we will show the supremacy of the latter over the former, which is consistent with the finding in most speech processing tasks \cite{RNNvsTransformer}. Our contributions in this work are three-fold:
\begin{itemize}
  \item We examine, through systematical objective and subjective evaluations, the TTS and ASR tasks for pretraining in seq2seq VC with different amounts of VC data.
  \item We analyze the hidden representation spaces of the learned models using different pretraining tasks.
  \item We compare the performance of RNN and Transformer architectures for seq2seq VC.
\end{itemize}

\section{Related Work}
\label{sec:related}

\subsection{Data deficiency in sequence-to-sequnece voice conversion}


In VC, it is common to limit the size of training data to around 5 or 10 minutes \cite{vcc2016, vcc2018}, wherein existing seq2seq VC literature, approximately 1 hour of data is usually used. Even with such amount of data, it is still required to resort to certain techniques to successfully train seq2seq VC models. These techniques can be categorized into the followings:

\noindent{\textbf{Extra module.}} Many have utilized an external ASR module pretrained on a large dataset during training and runtime. For example, the phonetic posteriorgram (PPG) extracted from ASR is a commonly used linguistic clue in VC \cite{VC-PPG}, and can be used as the only input \cite{S2S-FAC-VC, S2S-Taco-VC} or as an additional clue \cite{S2S-iFLYTEK-VC, S2S-Text-VC} in seq2seq VC. On the other hand, in \cite{S2S-parrotron-VC}, an external TTS system was used to generate artificial data from a large hand-transcribed corpus for training a many-to-one (normalization) VC model. The disadvantage of using external modules is that the performance depends on the extra module. The accuracy of the PPG and the quality of the TTS system can bound the performance of the final VC system.

\noindent{\textbf{Text label.}} Text labels provide strong supervision to ensure linguistic consistency. Methods utilizing such labels include multitask learning meaningful hidden representation \cite{S2S-Text-VC, S2S-parrotron-VC, S2S-NP-VC}, data augmentation \cite{S2S-Text-VC} or representations disentanglement \cite{S2S-NP-VC}. Yet, labeling errors and failed force alignments might cause potential performance degradation.

\noindent{\textbf{Regularization.}} As multitask learning and feature disentanglement can be seen as regularizations, some have also proposed to impose constraints on the model without any external resources. \cite{ATT-S2S-VC, CONV-S2S-VC} proposed the context preservation loss and the guided attention loss, and \cite{S2S-FAC-VC} proposed to use local attention to stabilize training. Nonetheless, such regularization often requires rigorous weight tuning.

Our proposed pretraining method is closest to the use of extra modules, except that the external resource is only used to obtain a prior for effective finetuning, instead of generating PPGs or artificial data. Also, our method needs neither text label of the VC data nor carefully designed regularization methods, yet can still achieve great data efficiency.

\subsection{Pretraining in speech processing}

Early applications of pretraining deep neural networks for speech processing mainly lied in ASR, with the main goal of speeding up optimization and reducing generalization error \cite{CD-DNN-HMM, DNN-ASR}. In recent years, inspired by the breakthrough in NLU, unsupervised or self-supervised speech representation learning utilizing massive, untranscribed speech data has become a popular research topic. As language modeling objectives have been widely employed for pretraining in NLU, finding a universally effective objective is still an active research area. Various objectives have been proposed, such as autoencoding \cite{wnae, PASE, VQVAE-code2spec} sometimes with an autoregressive model \cite{APC, MT-APC} or contrastive learning \cite{speaker-representation-MI, CPC, wav2vec, vq-wav2vec}. Nonetheless, different pretraining objectives lead to different representations, and an effective objective for VC is still unclear.

Different from the above mentioned unsupervised approaches, our method relies on \textit{supervised} pretraining with well-defined speech processing objectives. As we adopt popular speech processing tasks, large scale datasets can be assumed easily accessible thanks to the vastly growing community. We also expect that performance of pretraining would benefit from the rapid development of state-of-the-art models, thus improving the quality of the downstream VC task.

\section{Model architectures}
\label{sec:model}

Most existing seq2seq VC models are based on RNNs \cite{S2S-VC, ATT-S2S-VC, S2S-iFLYTEK-VC, S2S-Text-VC, S2S-NP-VC, S2S-FAC-VC, S2S-Hier-VC, S2S-parrotron-VC, S2S-Taco-VC}. In our prior work \cite{VTN}, we successfully applied the Transformer architecture \cite{transformer} to seq2seq VC. We first provide a unified model structure of seq2seq VC and describe the shared components.

%
%
%
%

\begin{figure}[t]
	\centering
	
	\begin{subfigure}[b]{0.48\textwidth}
		\centering
  		\includegraphics[width=0.6\textwidth]{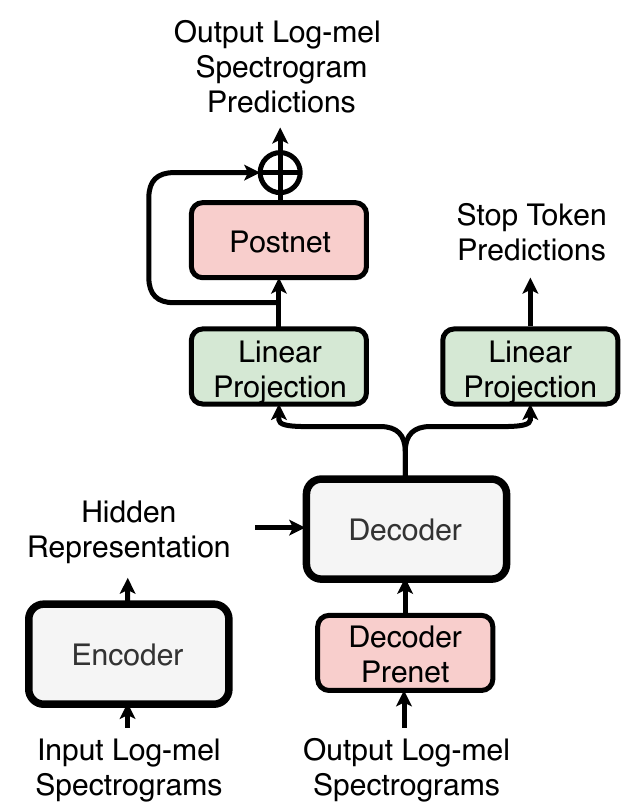}
		\caption{Illustration of the unified seq2seq VC model architecture and the shared components.}
   		\label{fig:seq2seq-vc-general}
	\end{subfigure}\\
	
	\vspace{0.5cm}
	
	\begin{subfigure}[b]{0.48\textwidth}
		\centering
	    \includegraphics[width=0.6\textwidth]{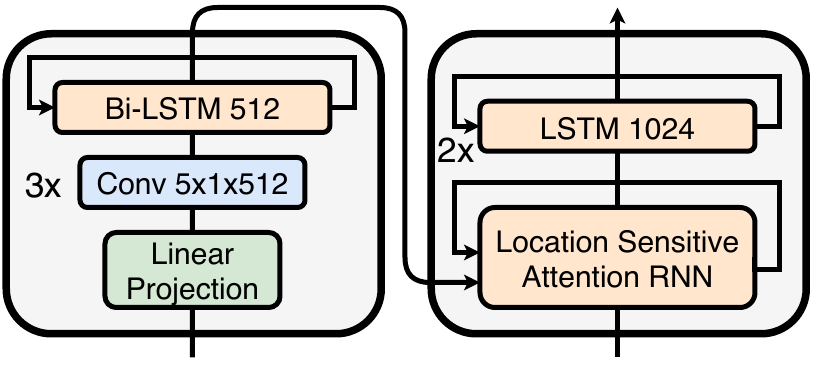}
		\caption{RNN-based encoder and decoder.}
   		\label{fig:seq2seq-vc-rnn}
   	\end{subfigure}\\
	
    \vspace{0.5cm}
	
	\begin{subfigure}[b]{0.48\textwidth}
		\centering
  		\includegraphics[width=0.8\textwidth]{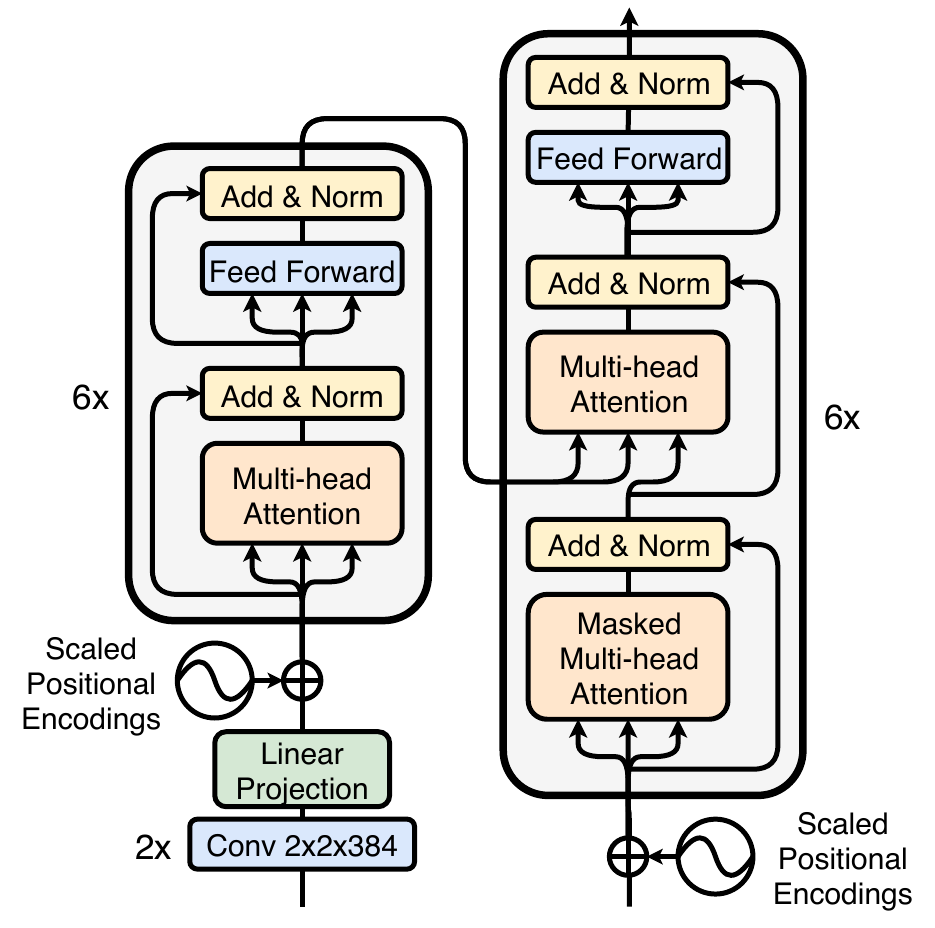}
		\caption{Transformer-based encoder and decoder.}
   		\label{fig:seq2seq-vc-VTN}
	\end{subfigure}
	
	\centering
	\caption{Model architectures that are used in this work.}
	\label{fig:architectures}
\end{figure}

\subsection{Unified seq2seq VC model structure}

Seq2seq models learn a mapping between a source feature sequence $X=\x_{1:n}=(\x_1, \cdots, \x_n)$ and a target feature sequence $Y=\y_{1:m}=(\y_1, \cdots, \y_m)$ which are often of different length, i.e, $n \neq m$. As with most seq2seq models, our seq2seq VC model also has an encoder\textemdash decoder structure \cite{S2S}, as dipicted in Figure~\ref{fig:seq2seq-vc-general}. The encoder ($\Enc$) first maps the input acoustic feature sequence $\x_{1:n}$ into a sequence of hidden representations:
\begin{align}
	H=\h_{1:n}= \Enc(\x_{1:n}).
\end{align}

The decoder ($\Dec$) is autoregressive, which means that when decoding the current output $\y_t$, in addition to the encoder output, i.e. the hidden representations $\h_{1:n}$, the previously generated features $\y_{1:t-1}$ are also considered:
\begin{align}
	\y_t = \Dec(	\h_{1:n}, \y_{1:t-1}).
\end{align}

Some extra components and techniques are adopted in our seq2seq VC model to improve performance and stabilize training, most of which are inspired by the success of modern seq2seq TTS models \cite{Taco, Taco2}.
\begin{itemize}
	\item A prenet containing 2 fully connected layers is added to the decoder, which serves as an information bottleneck essential for learning the autoregressive decoder. 
	\item A linear projection layer is used to project the decoder output to have the desired dimension. To learn when to stop decoding, a separate linear projection layer is used to predict a stopping probability, which can be used with a threshold to decide when to stop decoding during inference 
	\item To compensate for the missing context information in the autoregressive decoder, a five-layer CNN postnet is used to predict a residual that is added to the projected output.
	\item Introducing the reduction factor $r$ greatly helps speed up convergence and reduce training time and memory footprint. Specifically, at each decoding step, $r$ non-overlapping frames are predicted. Since adjacent speech frames are often correlated, this technique allows the decoder to correctly model the interaction with the hidden representation sequence.
\end{itemize}
The training objectives include an L1 and L2 loss, in combination with a weighted binary cross-entropy loss on the stop token prediction. The whole network is composed of neural networks and optimized via backpropagation.

\subsection{RNN-based model}
\label{ssec:seq2seq-vc-rnn}

Our RNN-based seq2seq VC model is based on the Tacotron2 TTS model \cite{Taco2} and resembles the work in \cite{ATT-S2S-VC}. The encoder first linearly projects the input log-mel spectrogram, followed by a stack of convolutional layers, batch normalization, and ReLU activations. The output of the final convolutional layer is then passed into a bi-directional LSTM layer to generate the hidden representations.

For each decoder output step, an attention mechanism \cite{S2S-NMT-Bah, S2S-NMT-Luong} is used to attend to different positions of the hidden representation sequence. First, a context vector $\vecc_t$ is calculated as a weighted sum of $\h_{1:n}$, where the weight is represented using an attention probability vector $\veca_t=(a^{(1)}_t, \cdots, a^{(n)}_t)$. Each attention probability $a^{(k)}_t$ can be thought of as the importance of the hidden representation $\h_k$ at the current time step. As in Tacotron2, we adopt the location-sensitive attention \cite{location-sensitive}, which takes cumulative attention weights from previous decoder time steps as an additional feature to encourage a forward consistency to prevent repeated or missed phonemes. The context vector is then concatenated with the prenet output and passed into a stacked uni-directional LSTM network to predict the $r$ output frames. 

For seq2seq speech synthesis models, the attention alignment is usually monotonic and linear, so a guided attention loss that encourages the attention matrix to be diagonal can speed up attention learning and convergence \cite{DC-GA-TTS, ATT-S2S-VC}. In addition, in \cite{ATT-S2S-VC}, a context preservation loss was applied to maintain linguistic consistency after conversion \cite{ATT-S2S-VC}.

\subsection{Transformer-based model}

Our Transformer-based seq2seq VC model, which we refer to as the Voice Transformer Network (VTN), is based on the Transformer architecture \cite{transformer}, which was originally designed for machine translation but also widely applied to other sequential modeling problems. There are several core components of the Transformer:

\noindent{\textbf{Multi-head attention (MHA) sublayer.}} The MHA layer is defined as:
\begin{align}
	\text{MHA}(Q,K,V) = [\head_1, \cdots, \head_{h}]W^{O},\\
	\head_i = \att(QW_i^Q, KW_i^K, VW_i^V),
\end{align}
where $Q$, $K$ and $V$ denote the input matrices that, following \cite{transformer}, are referred to as the query, key and value, respectively. MHA uses $h$ different, learned linear projections $W^Q, W^K, W^V$ to map the inputs to different \textit{heads}, and then perform the $\att$ operation in parallel. The outputs from all heads are concatenated and projected with $W^O$. As in \cite{transformer}, the $\att$ operation is implemented scaled dot-product attention is used:
\begin{equation}
	\att(Q,K,V) = \text{softmax}(\frac{QK^T}{\sqrt{d_{att}}})V,
\end{equation}
where $d_{att}$ is the attention dimension.

\noindent{\textbf{Position-wise feed-forward network (FFN) layer.}} The FFN layer is defined as:
\begin{align}
	FFN(\x) = \max(0, \x W^1 + b^1 )W^2 + b^2,
\end{align}
which is independently applied at each time step (position) with different parameters from layer to layer.

\noindent{\textbf{Layer normalization and residual connection.}} Around either of the above-mentioned sublayers, a residual connection followed by layer normalization \cite{layernorm} is employed. For input $X$ of a sublayer, the output is given as:
\begin{align}
	\text{LayerNorm}(X + \text{Sublayer}(X)).
\end{align}
Due to the residual connections, all sublayers have the same output dimension $\dmodel$.

\noindent{\textbf{Scaled positional encoding (SPE).}} In the original Transformer \cite{transformer}, since no recurrent relation is employed in the Transformer, to let the model be aware of information about the relative or absolute position of each element, the triangular (sinusoidal) positional encoding (PE) \cite{convs2s} is added to the inputs to the encoder and decoder. In this work, we adopt the SPE \cite{transformer-tts}, which is a generalized version of the original PE that scales the encodings with a trainable weight $\alpha$, so that they can adaptively fit the scales of the encoder and the decoder:
\begin{align}
	\text{SPE}(t)=
		\begin{cases}
			\alpha \cdot \sin(\frac{t}{10000^{\frac{2t}{\dmodel}}}), & \text{if }t \text{ is even,}\\
			\alpha \cdot \cos(\frac{t}{10000^{\frac{2t}{\dmodel}}}), & \text{if }t \text{ is odd.}
		\end{cases}
\end{align}

The encoder we adopt in this work resembles the one in \cite{transformer-asr}. First, the input acoustic feature sequence is downsampled in the time and frequency axes by a fraction of 4 using two convolutional layers with stride $2\times 2$. While the reduction of the memory footprint is a clear benefit, a hidden representation with a low sampling rate can not only speed up attention learning convergence due to easier attention calculation at each decoding step but also approximates phoneme-level or even character-level linguistic contents \cite{S2S-iFLYTEK-VC}. After linearly projecting to $\dmodel$-dimensions and adding the SPE, $L$ identical encoder layers are stacked to form the core of the encoder. Each encoder layer consists of an MHA sublayer and an FFN sublayer, followed by a residual connection and layer normalization. The MHA layers in the encoder are \textit{self-attention} layers since the queries, keys, and values are all from the output from the previous layer.

The decoder in this work is composed of the same number of $L$ identical decoder layers as in the encoder. In each decoder layer, the first sublayer is the so-called \textit{masked} self-attention MHA sublayer, where a mask is utilized such that at time step $t$, only vectors with time index up to and including $t$ can be accessed. This preserves the autoregressive property of the model. Then, an MHA sublayer uses the outputs from the previous layer as queries and $H$ as the keys and values, which ensembles the encoder\textemdash decoder attention in~\ref{ssec:seq2seq-vc-rnn}. Finally, an FFN sublayer is used, as in the encoder. Again, all sublayers are wrapped with a residual connection and layer normalization.

In addition to the L1, L2, and weighted binary cross-entropy losses, the guided attention loss is also applied. As pointed out in \cite{transformer-tts}, in Transformer-based speech synthesis, not all attention heads demonstrate diagonal alignments, so following \cite{RNNvsTransformer, espnet-tts}, the guided attention loss is applied to partial heads in partial decoder layers.

\begin{figure}[t]
  \centering
  \includegraphics[width=0.48\textwidth]{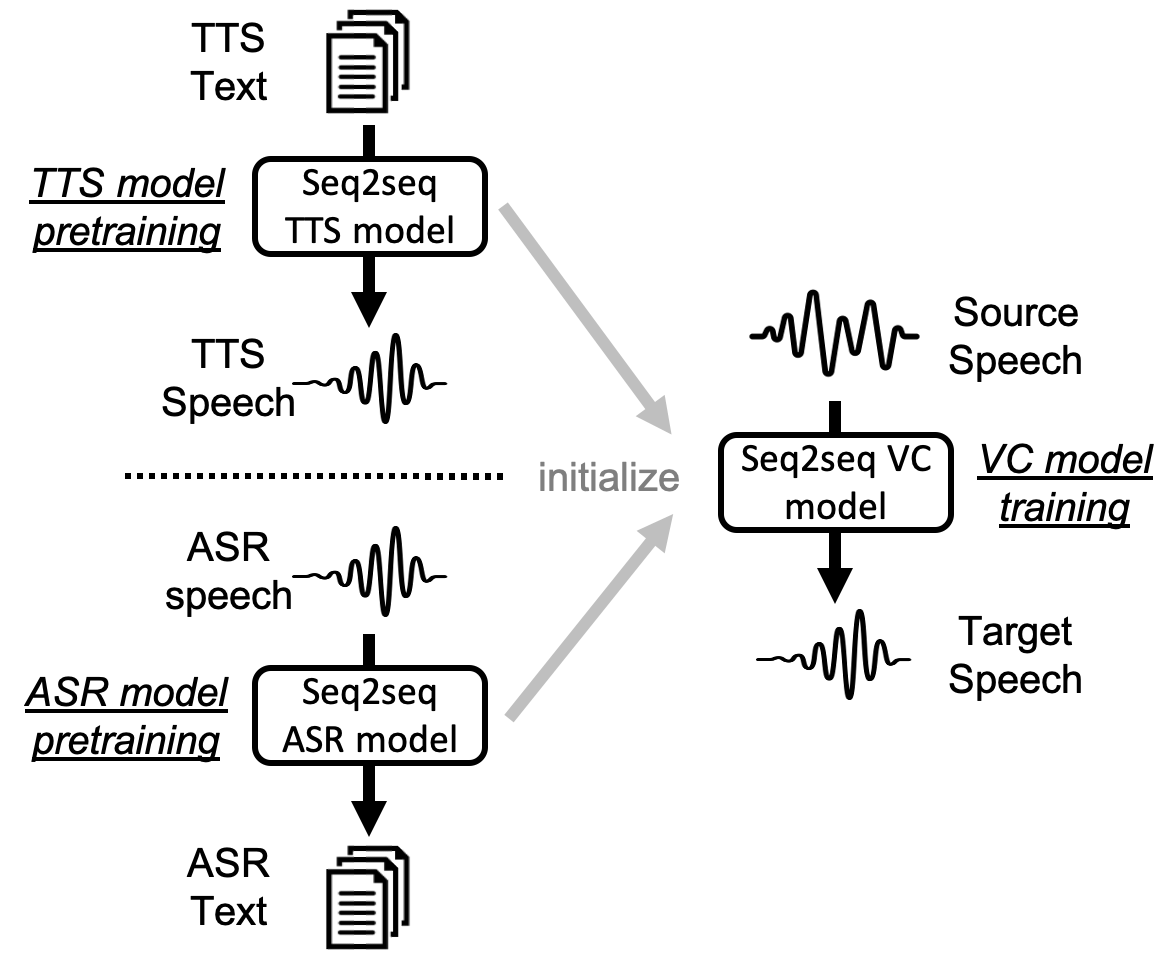}
  \centering
  \captionof{figure}{Illustration of the concept of pretraining from seq2seq TTS or ASR to seq2seq VC.}
  \label{fig:pt-for-seq2seq-vc}
\end{figure}

\begin{figure*}[t]
  \centering
  \includegraphics[width=\textwidth]{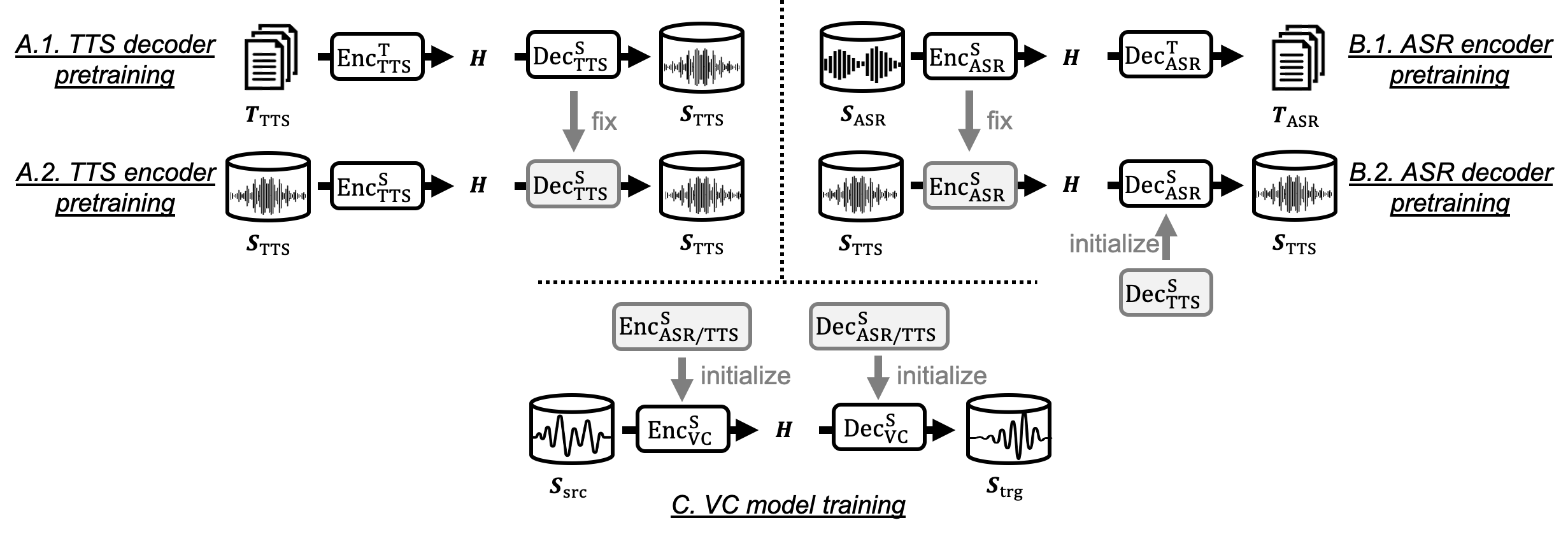}
  \centering
  \captionof{figure}{Diagram of the pretraining procedures for TTS and ASR. Top left: TTS pretraining. Top right: ASR pretraining. Bottom: VC model training.}
  \label{fig:pt-detail}
\end{figure*}

\section{Pretraining techniques}
\label{sec:pt}

In seq2seq models for speech applications, effective intermediate representations can facilitate correct attention learning that bridges the encoder and the decoder, thus crucial to success. By the definition of VC, it is natural to try to encode the linguistic contents of the source speech into the hidden representations so that they can be maintained. Thus, we conjecture that the core ability of successful seq2seq VC models is to generate and utilize high-fidelity hidden representations.

In theory, both TTS and ASR tasks aim to find a mapping between two modalities: speech and text. As speech signals contain all essential linguistic information, the hidden representation spaces induced by these two tasks should lie in the middle of the spectrum between speech and text. Thus, we conjecture that such space is desirable for seq2seq VC models, thus suitable for pretraining.

In this work, We extend the TTS-based pretraining technique previously proposed in \cite{VTN} to both TTS and ASR, as depicted Figure~\ref{fig:pt-for-seq2seq-vc}. Specifically, a two-stage training procedure is employed: in the first pretraining stage, a large-scale corpus is used to learn the initial seq2seq model parameters as a prior; then, in the second stage, the seq2seq VC model is initialized with the pretrained model parameters and trained with a relatively smaller VC dataset. The goal of this pretraining procedure is to provide fast, sample-efficient VC model learning, thus reducing the data size requirement and training time. In addition, this setup is highly flexible in that we do not require any of the speakers to be the same, nor any of the sentences between the pretraining corpus and the VC dataset to be parallel. 

Let the parallel VC dataset be $\DVC=\{\bm{S}_{\text{src}}, \bm{S}_{\text{trg}}\}$, where $\bm{S}_{\text{src}}, \bm{S}_{\text{trg}}$ denote the source, target speech, respectively. Our goal is to find a set of prior model parameters to train the final encoder $\Enc^{\text{S}}_{\text{VC}}$ and decoder $\Dec^{\text{S}}_{\text{VC}}$.

\subsection{TTS pretraining}
\label{ssec:tts-pt}

In this subsection we review the TTS pretraining technique \cite{VTN}. We assume that access to a large single-speaker TTS corpus $\DTTS=\{\bm{T}_{\text{TTS}}, \bm{S}_{\text{TTS}}\}$ is available, where $\bm{T}_{\text{TTS}}, \bm{S}_{\text{TTS}}$ denote the text and speech of the TTS speaker respectively. The pretraining can be broken down into two steps. 
\begin{itemize}
	\item[A.1] \textit{Decoder pretraining}: As in A.1 in Figure~\ref{fig:pt-detail}, the decoder is pretrained, on $\DTTS$, by training a conventional TTS model composed of a text encoder $\Enc^{\text{T}}_{\text{TTS}}$ and a speech decoder $\Dec^{\text{S}}_{\text{TTS}}$.
	\item[A.2] \textit{Encoder pretraining}: Then, as in A.2 in Figure~\ref{fig:pt-detail}, the encoder is pretrained, also on the same $\DTTS$, by training an autoencoder which takes $\bm{S}_{\text{TTS}}$ as input and output. The decoder here is the pretrained $\Dec^{\text{S}}_{\text{TTS}}$ and we fix the parameters so that they are not updated during training. The desired pretrained encoder $\Enc^{\text{S}}_{\text{TTS}}$ can then be obtained by minimizing the reconstruction loss.
\end{itemize}
The intuition of the encoder pretraining is to obtain an encoder capable of encoding acoustic features into hidden representations that are recognizable by the well pretrained decoder. Another interpretation is that the final pretrained encoder $\Enc^{\text{S}}_{\text{TTS}}$ tries to mimic the text encoder $\Enc^{\text{T}}_{\text{TTS}}$. In the first decoder pretraining step, since text itself contains pure linguistic information, the text encoder $\Enc^{\text{T}}_{\text{TTS}}$ is ensured to learn to encode an effective hidden representation that can be consumed by the decoder $\Dec^{\text{S}}_{\text{TTS}}$. Fixing the decoder in the encoder pretraining process, as a consequence, guarantees the encoder to behave similarly to the text encoder, which is to extract fine-grained, linguistic-information-rich representations.

\subsection{ASR pretraining}
\label{ssec:asr-pt}

In this subsection we describe how to extend the TTS pretraining technique in~\ref{ssec:tts-pt} to ASR. We assume that a large \textit{multi-speaker} ASR corpus $\DASR=\{\bm{S}_{\text{ASR}}, \bm{T}_{\text{ASR}}\}$ is available, where $\bm{S}_{\text{ASR}}, \bm{T}_{\text{ASR}}$ denote the speech and text data in $\DASR$, respectively. Similar to TTS pretraining, the ASR pretraining is again broken down into two steps. 
\begin{itemize}
	\item[B.1] Encoder pretraining: First, the \textit{encoder} is pretrained, on $\DASR$, by training a conventional ASR model consisting a speech encoder $\Enc^{\text{S}}_{\text{ASR}}$ and a text decoder $\Dec^{\text{T}}_{\text{ASR}}$, as in B.1 in Figure~\ref{fig:pt-detail}.
	\item[B.2] Decoder pretraining: Differently, the decoder pretraining is performed on $\DTTS$, rather on $\DASR$. This is because $\DASR$ is a multi-speaker corpus, but the VC model architecture in this work focuses on one-to-one VC, i.e. modeling the conversion between one source speaker and one target speaker, thus cannot model individual speaker characteristics. Again, the decoder pretraining uses $\bm{S}_{\text{TTS}}$ as input and output, and the encoder is the pretrained $\Enc^{\text{S}}_{\text{ASR}}$ and kept fixed during training. To speed up convergence, we initialize the decoder with the one obtained in TTS decoder pretraining, namely $\Dec^{\text{S}}_{\text{TTS}}$. The desired pretrained decoder $\Enc^{\text{S}}_{\text{ASR}}$ can then be obtained by minimizing the reconstruction loss. The decoder pretraining procedure is depicted in B.2 in Figure~\ref{fig:pt-detail}.
\end{itemize}

The intuition of the ASR decoder pretraining is different from that of the TTS encoder pretraining. The ASR speech encoder $\Enc^{\text{S}}_{\text{ASR}}$, trained with the ASR objective, should generate a compact hidden representation for decoding underlying linguistic contents. Such representations are believed to be easier to map to speech, thus suitable for pretraining the speech decoder $\Dec^{\text{S}}_{\text{ASR}}$.

\subsection{VC model training}

Finally, as in \cite{VTN}, $\DVC$ is used to train the desired VC models $\Enc^{\text{S}}_{\text{VC}}$ and $\Dec^{\text{S}}_{\text{VC}}$, with the encoder initialized with either $\Enc^{\text{S}}_{\text{TTS}}$ or $\Enc^{\text{S}}_{\text{ASR}}$, and the decoder with $\Dec^{\text{S}}_{\text{TTS}}$ or $\Dec^{\text{S}}_{\text{ASR}}$, respectively. As we will show later, the pretrained model parameters serve as a very good prior to adapt to the relatively scarce VC data, achieving significantly better conversion performance.

\section{Experimental evaluations}
\label{sec:experiments}

\subsection{Experimental settings}
\label{ssec:settings}

\subsubsection{Data}

 For $\DVC$, we conducted our experiments on the CMU ARCTIC database \cite{CMU-Arctic}, which contains parallel recordings of professional US English speakers sampled at 16 kHz. Data from four speakers were used: a male source speaker (\textit{bdl}) and a female source speaker (\textit{clb}), as well as a male target speaker (\textit{rms}) and a female target speaker (\textit{slt}). 100 utterances were selected for each validation and evaluation sets, and the remaining 932 utterances were used as training data. For $\DTTS$, we chose a US female English speaker (\textit{judy bieber}) from the M-AILABS speech dataset \cite{M-AILABS}. With the sampling rate also at 16 kHz, the training set contained 15,200 utterances, which were roughly 32 hours long. For $\DASR$, we used the LibriSpeech dataset \cite{librispeech} and pooled \textit{train-clean-100} and \textit{train-clean-360} together to get 460 hours of data from roughly 1170 speakers.

\subsubsection{Implementation}

The entire experiment was carried out on the open-source ESPnet toolkit \cite{espnet-tts, espnet}, including feature extraction, training and benchmarking. The official implementation has been made publicly available\footnote{\url{https://github.com/espnet/espnet/tree/master/egs/arctic/vc1}}, and since readers may access all the settings and configurations online, we omit the detailed hyperparameters here.
For the VC training, 80-dimensional mel filterbanks with 1024 FFT points and a 256 point frame shift was used as the acoustic features. We used the \textsc{LAMB} optimizer \cite{LAMB} and set the learning rate to 0.001.

\subsubsection{Waveform synthesis module}

We used the Parallel WaveGAN (PWG) \cite{parallel-wavegan}, which is a non-autoregressive variant of the WaveNet vocoder \cite{wavenet, sd-wnv} and enables parallel, faster than real-time waveform generation\footnote{We followed the open-source implementation at \url{https://github.com/kan-bayashi/ParallelWaveGAN}}. Since speaker-dependent neural vocoders outperform speaker-independent ones \cite{si-wnv}, we trained a speaker-dependent PWG conditioned on natural mel spectrograms, one for each target speaker. Note that we used the full training dataset, since the goal is to demonstrate the effects of various methods, so we did not train separate PWGs w.r.t. different training data sizes.

\subsubsection{Objective evaluation metrics}

We carried out two types of objective evaluations between the converted speech and the ground truth.
\begin{itemize}
	\item Mel cepstrum distortion (MCD): The MCD is a commonly used measure of spectral distortion in VC, which is based on mel-cepstral coefficients (MCCs). It is defined as: 
		\begin{equation}
		    MCD [dB] = \frac{10}{\log 10} \sqrt{2 \sum_{d=1}^K (mcc^{(c)}_d - mcc^{(t)}_d )^2},
		\end{equation}
		where $K$ is the dimension of the MCCs and $mcc^{(c)}_d$ and $mcc^{(t)}_d$ represent the $d$-th dimensional coefficient of the converted MCCs and the target MCCs, respectively. In practice, MCD is calculate in a utterance-wise manner. A dynamic time warping (DTW) based alignment is performed to find the corresponding frame pairs between the non-silent converted and target MCC sequences beforehand. We used the WORLD vocoder \cite{WORLD} for MCC extraction and silence frame decisions, and set $K=24$.
	\item Character/word error rate (CER/WER): The CER/WER is an underestimate of the intelligibility of the converted speech. The ASR engine is based on the Transformer architecture \cite{transformer-asr} and is trained using the LibriSpeech dataset \cite{librispeech}. The CER and WER for the ground-truth evaluation set were 0.9\% and 3.8\%, respectively.
\end{itemize}

\subsubsection{Subjective evaluation methods}

The following subjective evaluations were performed using the open-source toolkit \cite{p808-open-source} which implements the ITU-T Recommendation P.808 \cite{p808} for subjective speech quality assessment in the crowd using the Amazon Mechanical Turk (Mturk), and screens the obtained data for unreliable ratings. We recruited more than fifty listeners.\footnote{A demo web page with samples used for subjective evaluation is available at \url{ https://unilight.github.io/Publication-Demos/publications/vtn-taslp/index.html}}
\begin{itemize}
    \item The mean opinion score (MOS) test on naturalness: Subjects were asked to evaluate the naturalness of the converted and natural speech samples on a scale from 1 (completely unnatural) to 5 (completely natural).
    \item The VCC \cite{vcc2018} style test on similarity: This paradigm was adopted by the VCC organizing committee. Listeners were given a pair of speech utterances consisting of a natural speech sample from a target speaker and a converted speech sample. Then, they were asked to determine whether the pair of utterances can be produced by the same speaker, with 4-level confidence of their decision, i.e., sure or not sure.
\end{itemize}

\begin{table*}[t]
	\centering
	\captionsetup{justification=centering}
	\caption{Validation-set objective evaluation results of VTNs with no pretraining, TTS pretraining, ASR pretraining, and RNN-based models with no pretraining and TTS pretraining, which are trained on different conversion pairs and different sizes of data.}
	\centering
	\begin{tabular}{ c c c c | c c c | c c c | c c c }
		\toprule
		& & \multicolumn{2}{c}{Conversion pair} & \multicolumn{3}{c}{932 training utterances} & \multicolumn{3}{c}{250 training utterances} & \multicolumn{3}{c}{80 training utterances} \\
		\cmidrule(lr){5-7} \cmidrule(lr){8-10} \cmidrule(lr){11-13}
		Model type & Pretraining & Source & Target & MCD & CER & WER & MCD & CER & WER & MCD & CER & WER \\
		
		\midrule
		
		\multirow{15}{*}[-13pt]{VTN} & \multirow{5}{*}[-3pt]{None} & \multirow{2}{*}{clb(F)} & slt(F) & 6.60 & 12.4 & 20.2 & 7.43 & 29.2 & 42.3 & 8.23 & 65.3 & 87.6 \\
		& & & rms(M) & 6.83 & 21.4 & 33.0 & 7.83 & 53.2 & 73.3 & 8.68 & 71.8 & 94.8 \\
		& & \multirow{2}{*}{bdl(M)} & slt(F) & 7.33 & 23.1 & 33.7 & 8.31 & 52.9 & 75.9 & 8.74 & 73.9 & 95.7 \\ 
		& & & rms(M) & 7.37 & 28.4 & 43.3 & 8.30 & 56.2 & 78.4 & 9.14 & 79.0 & 102.8 \\
		\cmidrule(lr){3-13}
		& & \multicolumn{2}{c}{Average} & 7.03 & 21.3 & 32.6 & 7.97 & 47.9 & 67.5 & 8.70 & 72.5 & 95.2\\ 
		
		\cmidrule(lr){2-13}
		
		& \multirow{5}{*}[-3pt]{TTS} & \multirow{2}{*}{clb(F)} & slt(F) & 6.02 & 5.5 & 9.1 & 6.41 & 5.2 & 9.7 & 6.66 & 10.4 & 14.7 \\
		& & & rms(M) & 6.22 & 6.8 & 11.9 & 6.75 & 12.8 & 21.3 & 6.94 & 12.5 & 22.0 \\
		& & \multirow{2}{*}{bdl(M)} & slt(F) & 6.33 & 5.0 & 7.6 & 6.71 & 4.8 & 8.1 & 7.07 & 9.7 & 13.6 \\ 
		& & & rms(M) & 6.69 & 7.3 & 12.7 & 7.13 & 11.3 & 18.0 & 7.39 & 17.2 & 26.2 \\
		\cmidrule(lr){3-13}
		& & \multicolumn{2}{c}{Average} & 6.32 & 6.2 & 10.3 & 6.75 & 8.5 & 14.3 & 7.02 & 12.5 & 19.1 \\ 
		
		\cmidrule(lr){2-13}
		
		& \multirow{5}{*}[-3pt]{ASR} & \multirow{2}{*}{clb(F)} & slt(F) & 6.11 & 4.8 & 10.9 & 6.84 & 15.9 & 26.0 & 8.28 & 72.1 & 97.6 \\
		& & & rms(M) & 6.22 & 8.1 & 16.0 & 7.08 & 27.2 & 43.2 & 7.93 & 60.2 & 86.2\\
		& & \multirow{2}{*}{bdl(M)} & slt(F) & 6.50 & 5.7 & 11.1 & 7.33 & 26.1	 & 39.8 & 8.18 & 58.2 & 80.7 \\ 
		& & & rms(M) & 6.68 & 9.1 & 15.6 & 7.58 & 32.9 & 51.6 & 8.22 & 59.7 & 82.9 \\
		\cmidrule(lr){3-13}
		& & \multicolumn{2}{c}{Average} & 6.38 & 6.9 & 13.4 & 7.21 & 25.5 & 40.2 & 8.15 & 62.6 & 86.9 \\ 
		
		\midrule
		
		\multirow{10}{*}[-7pt]{RNN} & \multirow{4}{*}[-3pt]{None} & \multirow{2}{*}{clb(F)} & slt(F) & 6.77 & 7.1 & 12.1 & 7.29 & 15.4 & 24.0 & 7.76 & 38.6 & 56.8\\
		& & & rms(M) & 6.80 & 11.6 & 19.7 & 7.49 & 24.7 & 38.0 & 7.98 & 48.9 & 68.7 \\
		& & \multirow{2}{*}{bdl(M)} & slt(F) & 7.45 & 23.4 & 32.6 & 8.06 & 37.1 & 54.4 & 8.44 & 65.6 & 93.8 \\ 
		& & & rms(M) & 7.62 & 20.0 & 32.4 & 8.25 & 47.2 & 90.2 & 8.52 & 59.7 & 81.5 \\
		\cmidrule(lr){3-13}
		& & \multicolumn{2}{c}{Average} & 7.16 & 15.5 & 24.2 & 7.77 & 31.1 & 51.7 & 8.18 & 53.2 & 75.2 \\ 
		
		\cmidrule(lr){2-13}
		
		& \multirow{4}{*}[-3pt]{TTS} & \multirow{2}{*}{clb(F)} & slt(F) & 6.29 & 5.6 & 10.1 & 6.63 & 7.4 & 12.7 & 6.92 & 14.0 & 22.2 \\
		& & & rms(M) & 6.35 & 8.3 & 16.1 & 6.88 & 17.0 & 27.7 & 7.08 & 29.0 & 44.0 \\
		& & \multirow{2}{*}{bdl(M)} & slt(F) & 6.74 & 8.2 & 13.9 & 7.08 & 11.3 & 19.8 & 7.46 & 16.3 & 23.8\\ 
		& & & rms(M) & 6.97 & 15.1 & 26.3 & 7.39 & 21.1 & 32.4 & 7.57 & 25.4 & 39.6 \\
		\cmidrule(lr){3-13}
		& & \multicolumn{2}{c}{Average} & 6.59 & 9.3 & 16.6 & 7.00 & 14.2 & 23.2 & 7.26 & 21.2 & 32.4 \\ 
		
		\bottomrule
	\end{tabular}
	\label{tab:obj-eval}
\end{table*}

\begin{table*}[t]
	\centering
	\captionsetup{justification=centering}
	\caption{ASR-based recognition results of VTN converted samples from the evaluation set of the clb-slt conversion pair. The errors are in uppercase.}
	\centering
	\begin{tabular}{ c c c }
		\toprule
		Description & Training data size & Recognition result \\
		\midrule
		Ground truth & - & the history of the eighteenth century is written ernest prompted \\
		\midrule
		\multirow{3}{*}{TTS pretraining} & 932 & the history of the eighteenth century is written IN IS prompted TO TO \\
		& 250 & the history of the eighteenth century is written IN HIS    PROMPTER \\
		& 80 & the history of the eighteenth century is written ON HIS    PROMPT \\
		\midrule
		\multirow{3}{*}{ASR pretraining} & 932 &  the history of the eighteenth century is written EARNEST prompted \\
		& 250 &  the history of the eighteenth CENTURY'S RADIANCE prompted \\
		& 80 & IT DISTURBED the DAY TO HIMSELF TO REJOIN HIM IN NORTH'S LIBRARY \\
		\bottomrule
	\end{tabular}
	\label{tab:recognition-samples}
\end{table*}

\begin{table*}[t]
	\centering
	\captionsetup{justification=centering}
	\caption{Evaluation-set naturalness and similarity subjective evaluation results of VTNs with no pretraining, TTS pretraining, ASR pretraining, and RNN-based models with no pretraining and TTS pretraining, which are averaged over all conversion pairs and different sizes of data.}
	\centering
	\begin{tabular}{ c c c c c c }
		\toprule
		& &  \multicolumn{2}{c}{932 training utterances} & \multicolumn{2}{c}{80 training utterances}  \\
		\cmidrule(lr){3-4} \cmidrule(lr){5-6}
		Model & Pretraining & Naturalness & Similarity & Naturalness & Similarity \\
		
		\midrule
		
		\multicolumn{2}{c}{Analysis-synthesis} & 4.45 $\pm$ 0.14 & - & - & - \\
		
		\midrule
		
		\multirow{3}{*}{VTN} & None & 3.19 $\pm$ 0.23 & 61\% $\pm$ 14\% & 1.96 $\pm$ 0.16 & 44\% $\pm$ 13\%  \\
		& TTS & 4.34 $\pm$ 0.15 & 80\% $\pm$ 11\% & 4.11 $\pm$ 0.09 & 68\% $\pm$ 8\%  \\
		& ASR & 4.25 $\pm$ 0.16 & 77\% $\pm$ 10\% & 3.38 $\pm$ 0.20 & 53\% $\pm$ 12\%  \\
		
		\midrule
		
		\multirow{2}{*}{RNN} & None & 2.33 $\pm$ 0.20 & 40\% $\pm$ 12\% & 1.57 $\pm$ 0.14 & 33\% $\pm$ 15\%  \\
		& TTS & 3.91 $\pm$ 0.19 & 68\% $\pm$ 13\% & 3.71 $\pm$ 0.09 & 58\% $\pm$ 10\%  \\

		\bottomrule
	\end{tabular}
	\label{tab:sub-eval}
\end{table*}

\subsection{Effectiveness of TTS pretraining on RNN and Transformer based models}

First, we show that TTS pretraining is a technique effective on not only VTN but also RNN-based seq2seq VC models. The objective results are in Table~\ref{tab:obj-eval}. First, without pretraining, both VTN and RNN could not stay robust against the reduction of training data. The performance dropped dramatically with the reduction of training data, where a similar trend was also reported in \cite{S2S-NP-VC}. This identifies the data efficiency problem of seq2seq VC. By incorporating TTS pretraining, both VTN and RNN exhibited a significant improvement in all objective measures, where the effectiveness was robust against the reduction in the size of training data. With only 80 utterances, both models can achieve comparable performance to that of using the full training dataset, wherein the case of VTN, the intelligibility is even better.

The subjective results are in Table~\ref{tab:sub-eval}. Without pretraining, the VTN and RNN suffered from about 1.2 and 0.8 MOS points drop when the training data reduces from 932 to 80 utterances. On the other hand, with TTS pretraining applied, the naturalness of VTN and RNN improved by more than 1 point with 932 utterances and more than 2 points with 80 utterances. Moreover, when the training data reduces, there was only a very limited performance drop. These results demonstrate the effectiveness of the TTS pretraining technique.

\subsection{Comparison of TTS and ASR pretraining}

Next, we compare the effectiveness of TTS and ASR pretraining. From Tables~\ref{tab:obj-eval} and~\ref{tab:sub-eval}, with the full training set, ASR pretraining could bring almost the same amount of improvement compared to TTS pretraining. However, as the size of training data reduces, the performance of the ASR pretrained model dropped significantly, showing that ASR pretraining lacks the robustness essential for practical VC.

To investigate the failure of ASR pretrained models against limited training data, we chose one sentence from the evaluation set and show the ASR results of the converted samples using TTS and ASR pretrained VTNs in Table~\ref{tab:recognition-samples}. Although TTS pretraining could not ensure complete linguistic consistency, the errors were minor and possibly due to the imperfect ASR engine used for evaluation, thus the result seems reasonable. On the other hand, the recognition result of the ASR pretrained model with 80 that had no connection to the source sentence. We conclude that linguistic consistency is poorly maintained under the limited data scenario using ASR pretraining.

\subsection{Comparison of RNN and Transformer based models}

In \cite{VTN}, the VTN was shown to outperform the RNN-based model \cite{ATT-S2S-VC}, while it was not clear whether the improvement came from different model architectures or the pretraining technique. To make a clearer comparison, we applied TTS pretraining to both VTNs and RNNs. From Tables~\ref{tab:obj-eval} and~\ref{tab:sub-eval}, it was shown that without TTS pretraining, VTNs performed worse than RNNs in terms of intelligibility measures but better in terms of subjective measures. This is possibly because that a more complex model like VTN is capable of generating better-sounding voices while being more prone to overfitting since it lacks attention regularizations such as the location-sensitive location, as suggested in \cite{espnet-tts}. As we applied TTS pretraining to both VTN and RNN, it could be clearly observed that VTNs outperformed RNNs in terms of all objective measures and subjective scores. 

\begin{figure}[t]
	\centering
	\begin{subfigure}{.25\textwidth}
	    \centering
	    \includegraphics[width=\textwidth]{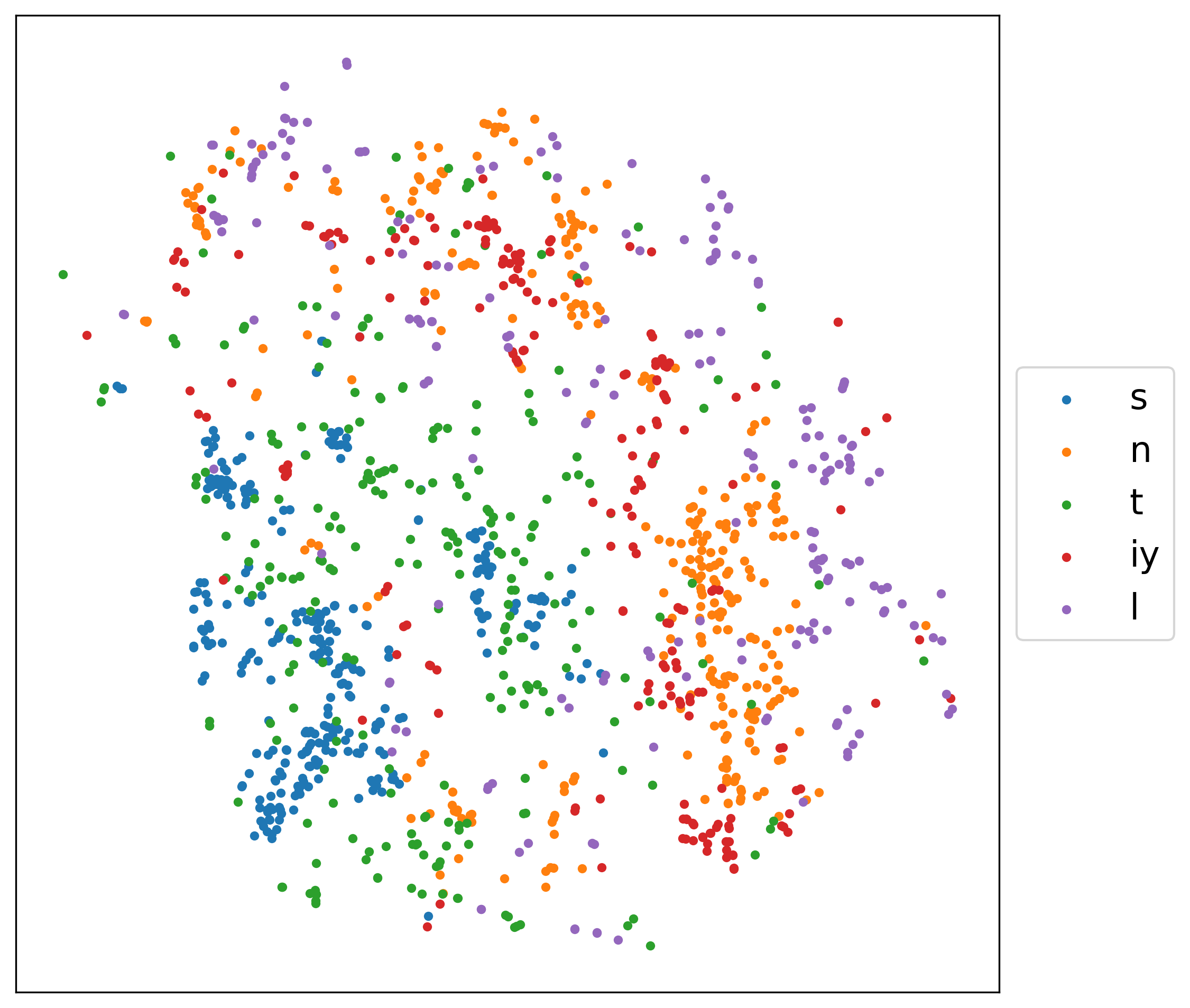}
	    \caption{No pretraining (932)}
   		\label{fig:vis-vtn-nopt-n932}
	\end{subfigure}%
	\begin{subfigure}{.25\textwidth}
	    \centering
		\includegraphics[width=\textwidth]{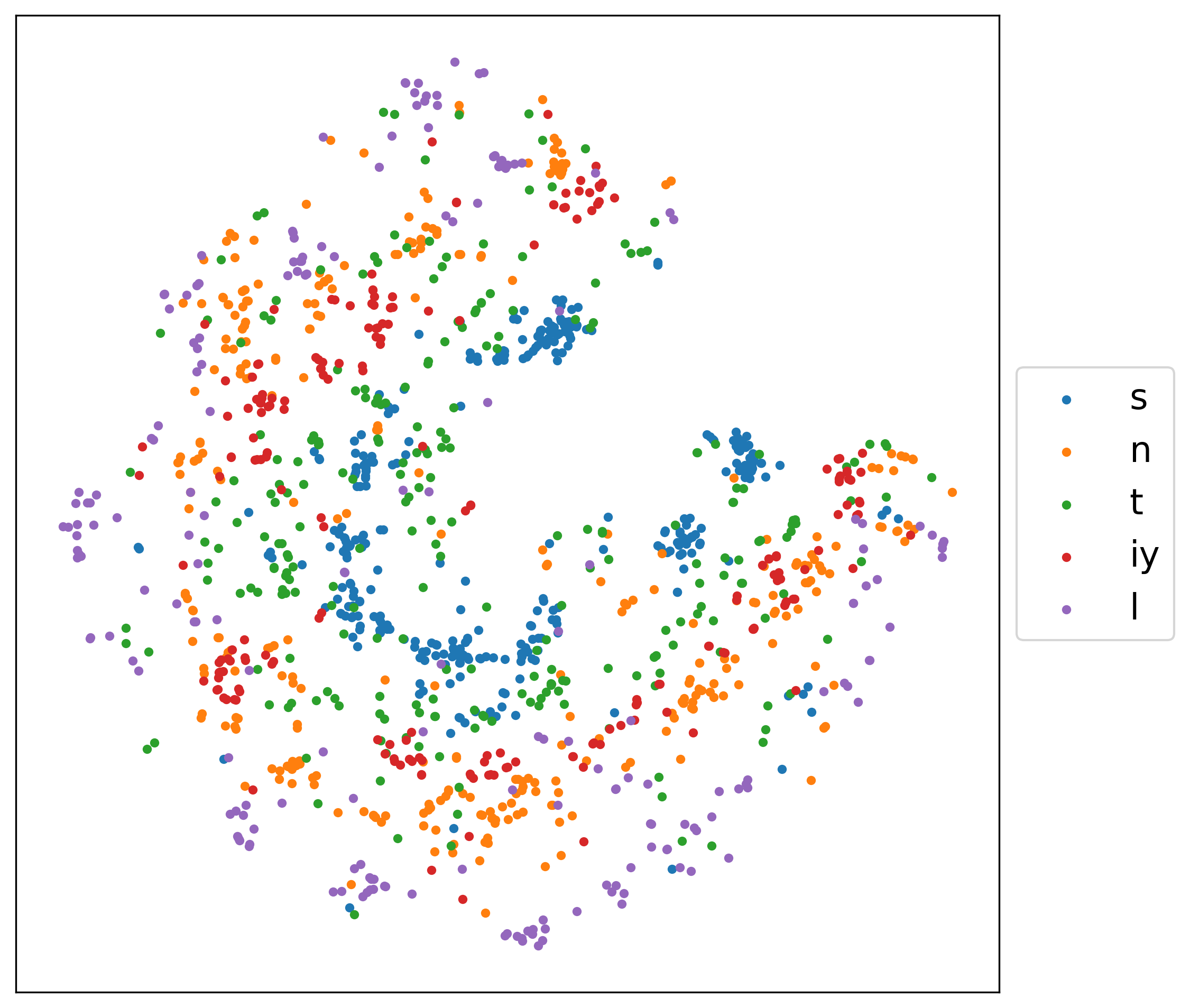}
		\caption{No pretraining (80)}
   		\label{fig:vis-vtn-nopt-n80}
	\end{subfigure}
	
	\begin{subfigure}{.25\textwidth}
	    \centering
	    \includegraphics[width=\textwidth]{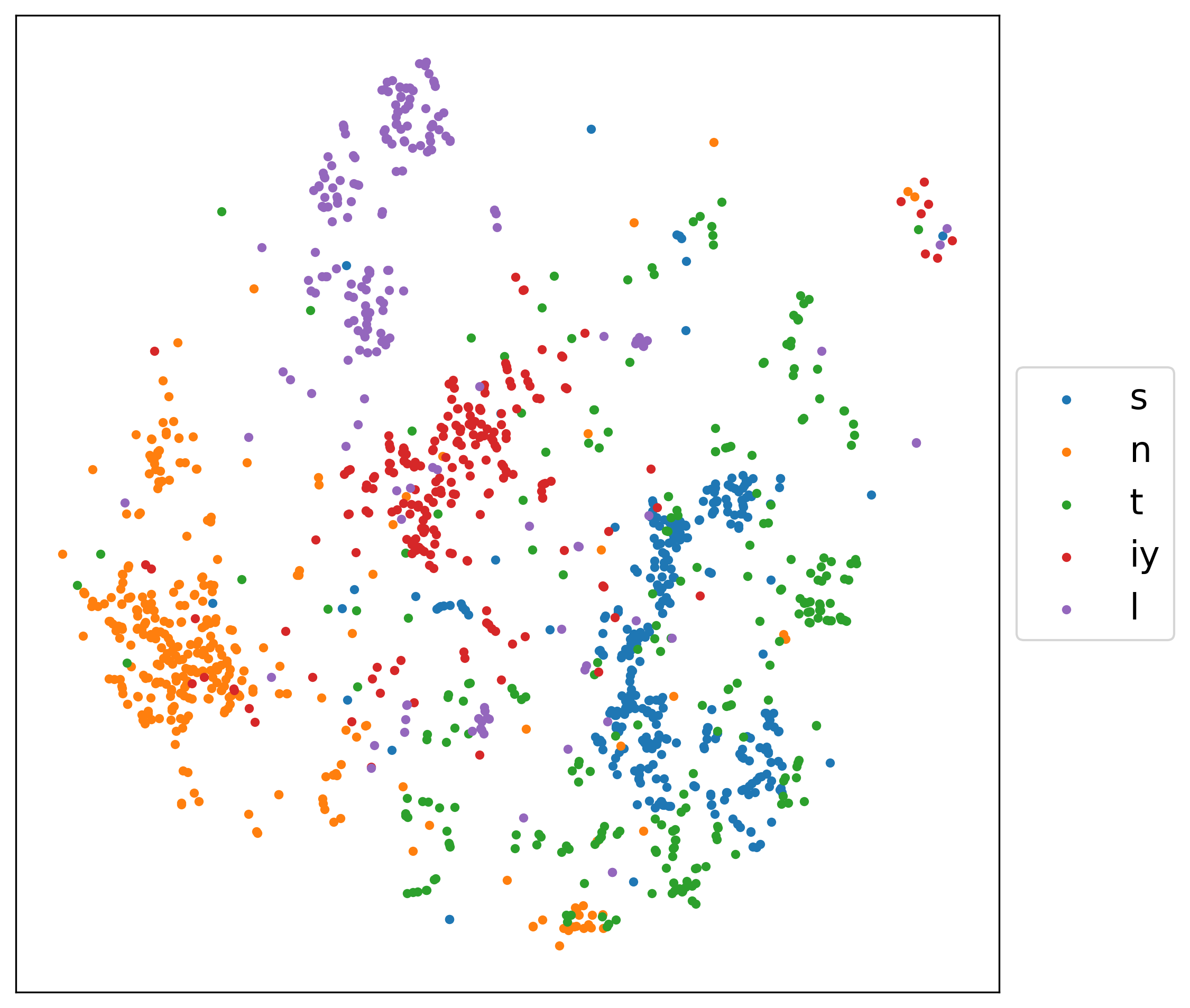}
	    \caption{TTS pretraining (932)}
   		\label{fig:vis-vtn-tts-pt-n932}
	\end{subfigure}%
	\begin{subfigure}{.25\textwidth}
	    \centering
		\includegraphics[width=\textwidth]{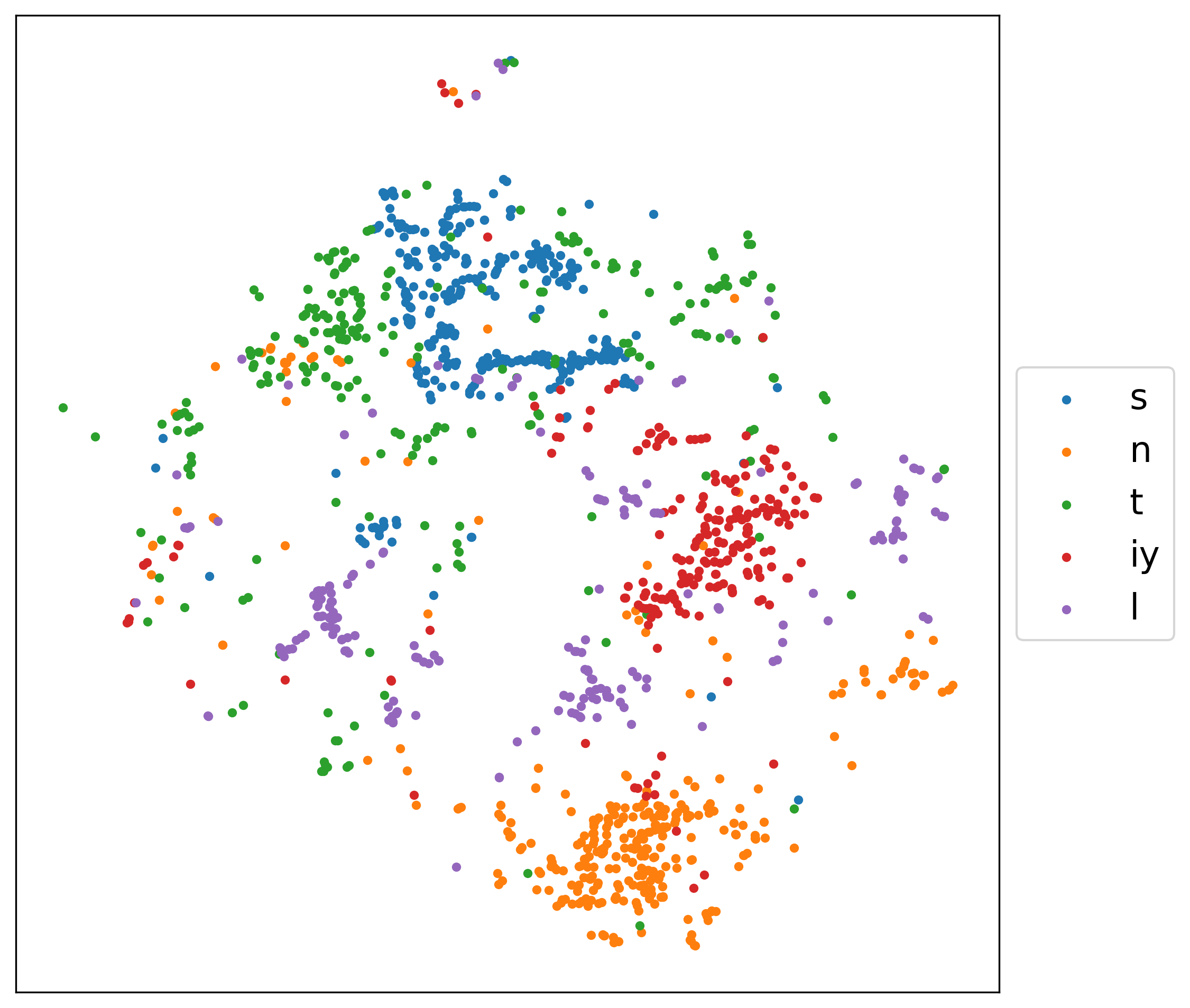}
		\caption{TTS pretraining (80)}
   		\label{fig:vis-vtn-tts-pt-n80}
	\end{subfigure}
	
	\begin{subfigure}{.25\textwidth}
	    \centering
	    \includegraphics[width=\textwidth]{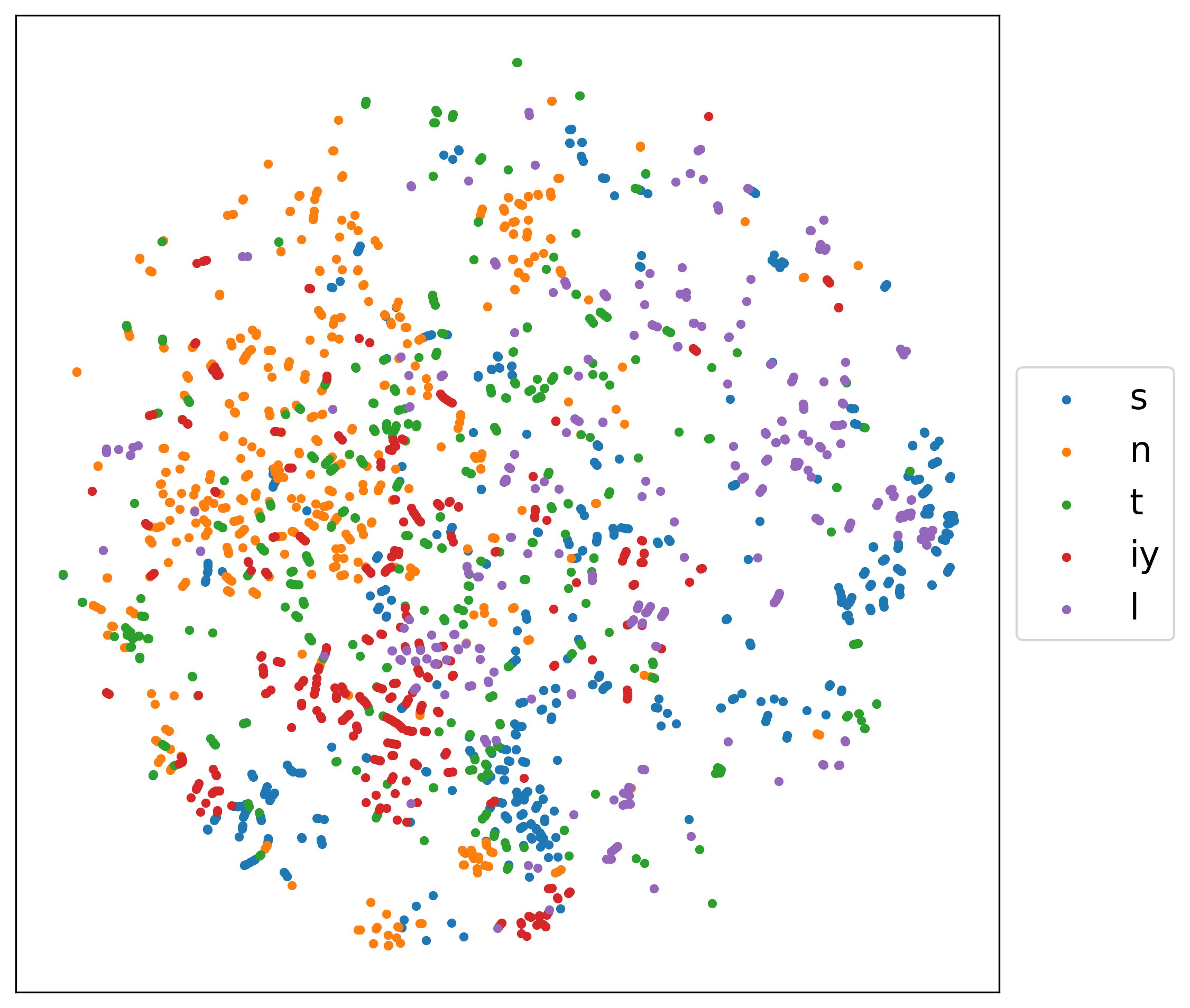}
	    \caption{ASR pretraining (932)}
   		\label{fig:vis-vtn-asr-pt-n932}
	\end{subfigure}%
	\begin{subfigure}{.25\textwidth}
	    \centering
	    \includegraphics[width=\textwidth]{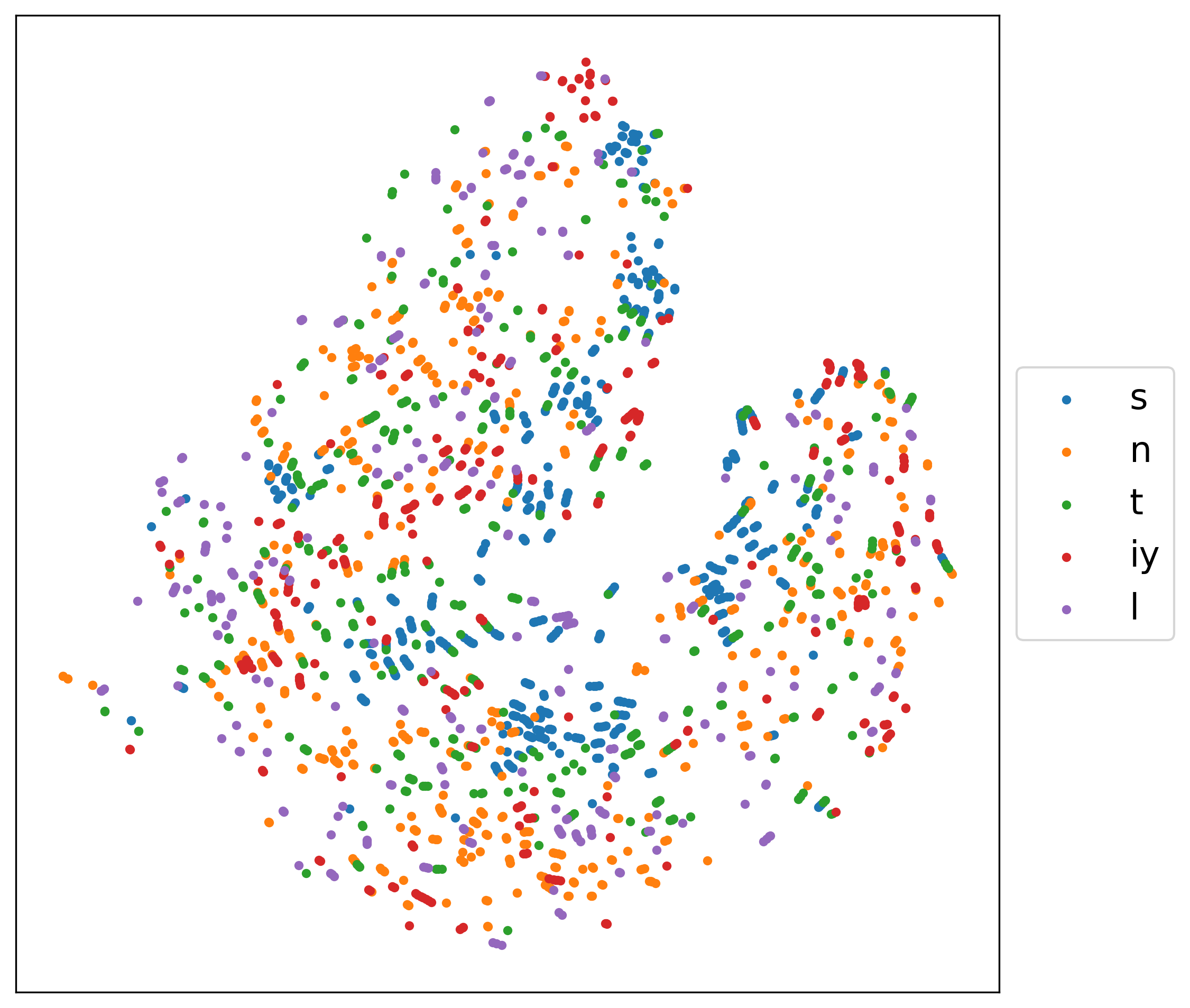}
	    \caption{ASR pretraining (80)}
   		\label{fig:vis-vtn-asr-pt-n80}
	\end{subfigure}
	
	\centering
	\caption{Visualizations of hidden representations extracted from VTNs with no pretraining, TTS pretraining, and ASR pretraining. The validation set from \textit{clb} was used. The numbers in the parenthesis indicate the number of training utterances.}
	\label{fig:visualization}
\end{figure}

\subsection{Visualizing the hidden representation space}

In Section~\ref{ssec:tts-pt}, we suspected that applying the TTS pretraining technique results in an encoder that can extract linguistic-information-rich representation. To demonstrate this tendency, we extracted the hidden representations with the trained encoders using the validation set from the \textit{clb} speaker as input, and visualized them using the t-SNE method \cite{tsne}. We used the phoneme labels that come with the CMU ARCTIC dataset as ground truth and colored the 5 most common phonemes and their corresponding hidden representations to simplify the plots.

The resulting plots are shown in Figure~\ref{fig:visualization}. It could be clearly observed that compared to no pretraining, the hidden representation spaces learned from TTS pretraining demonstrated a strong degree of clustering effect, where points correspond to the same phoneme were close to each other. This tendency was consistent in the cases of both 932 and 80 training utterances.
On the other hand, ASR pretraining yielded a much scatter hidden representation space even with 932 training utterances.

This analysis suggests that the TTS pretraining technique can result in a more discretized representation space, which matches our initial assumption. We may further conclude that, by looking together with the objective and subjective results in Tables~\ref{tab:obj-eval} and~\ref{tab:sub-eval}, the degree of clustering effect somehow reflect the goodness of the hidden representations for seq2seq VC.

\section{Conclusions}
\label{sec:conclusion}

In this work, we evaluated the pretraining techniques for addressing the problem of data efficiency in seq2seq VC. Specifically, a unified, two-stage training strategy that first pretrains the decoder and the encoder subsequently followed by initializing the VC model with the pretrained model parameters was proposed. ASR and TTS were chosen as source tasks to transfer knowledge from, and the RNN and VTN architectures were implemented. Through objective and subjective evaluations, it was shown that the TTS pretraining strategy can greatly improve the performance in terms of speech intelligibility and quality when applied to both RNNs and VTNs, and the performance could stay without significant degradation even with limited training data. As for ASR pretraining, the robustness was not so good with the reduction of training data size. Also, VTNs performed inferior to RNNs without pretraining but superior with TTS pretraining. The visualization experiment suggested that the TTS pretraining could learn a linguistic-information-rich hidden representation space while the ASR pretraining lacks such ability, which lets us imagine what an ideal hidden representation space would be like.

In the future, we plan to extend our pretraining technique to more flexible training conditions, such as many-to-many \cite{m2m-vtn} or nonparallel training \cite{S2S-NP-VC}.




\bibliographystyle{IEEEtran}
\bibliography{ref}






\end{document}